# Heavy-light baryonic mass splittings from the lattice

C. Alexandrou [a], A. Borrelli [b], S. Güsken [c], F. Jegerlehner[b], K. Schilling[c] [d], G. Siegert[c] and R. Sommer[d]

[a]*Department of Natural Sciences, University of Cyprus, Nicosia, Cyprus*

[b]*Paul Scherrer Institute, CH-5232 Villigen PSI, Switzerland*

[c]*Physics Department, University of Wuppertal D-42097 Wuppertal, Germany*

[d]*CERN, Theory Division, CH-1211 Geneva-23, Switzerland*

We present here lattice estimates of the heavy-light baryon $\Lambda_b$, obtained using propagating heavy quarks. Our result is: $M_{\Lambda_b} = 5.728 \pm 0.144 \pm 0.018$ Gev, after extrapolation to the continuum limit and in the quenched approximation.

## 1. Introduction

Since many years, lattice QCD is being applied to the computation of hadronic spectroscopy, with considerable success [1]. In connection with the problem of CP violation, much attention has been devoted to the study of heavy-light quark systems. In particular, heavy-light mesons like the $D$ and the $B$ [2] have been extensively studied.

In more recent times, studies of heavy-light baryon states have begun. The first results were exploratory computations of heavy-light baryon masses, in the infinite mass limit of the heavy quark [3,4].

Here, we present the results of a recent lattice study of heavy-light baryonic systems, with finite heavy quark masses, performed in the quenched approximation, using Wilson fermions [5].

We estimated the masses of the baryons $\Lambda_b$ (composed of one $b-$quark and two light ones) and $\Xi_b$ (two $b-$quarks and a light one).

At this same conference, some preliminary results on heavy-light baryon masses were presented by members of the UKQCD collaboration, both for propagating [6] and for static [7] heavy quarks. Their predictions turned out to be in agreement with ours, as will be seen later.

## 2. Computation Strategy

As is well known, direct simulation of $b-$quarks on present-day lattices would lead to systematically wrong results, due to the large $b-$quark mass. To avoid such systematical errors, we performed our computation using some precautions:

- we did not go too near to the $b$ mass, but only computed up to approximately twice the $c-$quark mass, and then extrapolated to the $b-$quark;
- we evaluated splittings with respect to heavy-light meson states, rather than directly the baryon masses; the splittings were chosen so that the heavy quark mass dependence would cancel in the infinite mass limit;
- we performed simulations on three different lattices, monitoring thus the $a$ dependence of the results.

## 3. Operators and smearing

We evaluated on the lattice the baryonic heavy-light-light and heavy-heavy-light correlators $C_\Lambda$ and $C_\Xi$ [4], corresponding to the $\Lambda$ and $\Xi$ baryons. Done that, we perfomed a fit of the ratios of those correlators to the pseudoscalar meson correlator $C_P$ [3]

$$R_\Lambda(t) = \frac{C_\Lambda(t)}{C_P(t)} \to A e^{-\Delta_\Lambda t} \qquad (1)$$

| $N_S$ | $N_T$ | number of conf. | $\beta$ | $a\, m_\rho$ |
|---|---|---|---|---|
| 8 | 24 | 175 | 5.74 | $0.542 \pm 0.014$ |
| 10 | 24 | 213 | 5.74 | |
| 12 | 24 | 113 | 5.74 | |
| 12 | 36 | 204 | 6.00 | $0.355 \pm 0.016$ |
| 18 | 48 | 67 | 6.26 | $0.260 \pm 0.014$ |

Tab.1: Lattice parameters

$$R_\Xi(t) = \frac{C_\Xi(t)}{C_P(t)^2} \to A e^{-\Delta_\Xi t} \qquad (2)$$

this yielded the values of the mass splittings:

$$\Delta_\Lambda = M_{\Lambda_b} - M_B \quad \text{and} \quad \Delta_\Xi = M_{\Xi_b} - 2M_B, \qquad (3)$$

in which the heavy quark mass dependence cancels in the infinite mass limit.

To obtain a good overlap of our matrix elements with the ground state, smearing [8] was applied to the light quark source. We made use in this computation of exactly the same smearing wave functions used in ref.[3], and, for almost all $(\kappa_l, \kappa_h)$ local mass channels, we obtained quite distinct plateaus.

### 4. Lattice parameters and finite volume effects

The simulations were performed for various values of $\kappa_l$ and $\kappa_h$ on the lattices whose parameters are reported in table 1.

The results from the $8^3 \times 24$, $12^3 \times 36$ and $18^3 \times 48$ lattices, at $\beta = 5.74$, 6.0, and 6.26, were used to compare between different $a$ values, check that the $a$ dependence is actually small, and perform the continuum limit.

At the same time, to check the dependence of the results on the lattice dimension, we compared the results from the three lattices at $\beta = 6.0$, with different volumes [9]. We noted that changing the lattice volume causes, as expected, only a small deviation of about 4%: this we considered as a systematic error in our final results.

### 5. The $\Delta_{\Lambda/\Xi}$ vs $\frac{1}{M_P}$ dependence and the continuum limit

For each lattice, we computed the mass splittings for five $\kappa_h$ values and three $\kappa_l$ ones; the light quark mass was then extrapolated to both the chiral limit and the strange quark mass (for details on this extrapolation, see [3]).

In ref. [9] the values of the pseudoscalar meson state $M_P$ for corresponding parameters had been reported, and we used them to plot the mass splitting $\Delta_{\Lambda/\Xi}$ vs $\frac{1}{M_P}$. Not all of the $\kappa_h$ values have in the end been used in the analysis: the lowest two for each lattice yielded pseudoscalar meson masses larger than $\sim 1.2$ in lattice units, and were therefore discarded to avoid systematical errors.

The remaining data were converted into physical units and plotted. To set the lattice scale, the value of the $\rho$-meson mass was used: a natural choice, for evaluating masses; in this way, systematical effects cancel, and we obtain a relatively small $a$ dependence.

In fig.1, the values of $\Delta_\Lambda$ with chiral light quarks are shown. The data exhibit a clear linear dependence on $\frac{1}{M_P}$,

$$\Delta_\Lambda = c_0 + c_1 \frac{1}{M_P}. \qquad (4)$$

As for finite $a$ effects, they appear to be quite small; neglecting them, by fitting directly all nine points with (4) (line in fig.1), and extrapolating then to $M_P = M_B = 5.28$ GeV, we obtain $\Delta_\Lambda^0 = 431 \pm 28$ MeV.

To take into account finite $a$ effects, we choose a set of values of $\frac{1}{M_P}$, in a region where distortions due to the heavy mass should not be present, and interpolate the lattice results from each fixed $\beta$ value to those $M_P$ values. This enables us to compare $\Delta_\Lambda$ at different values of $a$. For each $\frac{1}{M_P}$ value, we then assume a linear $a$ dependence for $\Delta_\Lambda$, extrapolate to the continuum limit, and fit the resulting set of $\Delta_\Lambda^{\text{cont}}$ values to (4). Finally, extrapolating to the $b$-quark mass, we obtain the result

$$\Delta_{\Lambda_b} = 458 \pm 144 \pm 18 \text{ MeV}, \qquad (5)$$

where the first error is purely statistical, and the second is the systematic error due to finite volume effects.

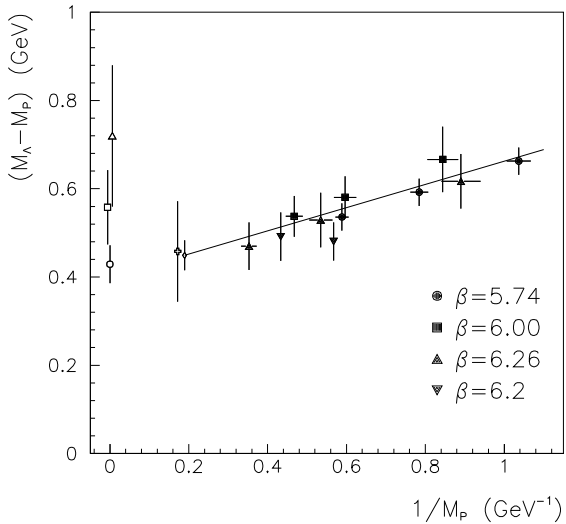

**Fig.1:** Our nine data points, at $\beta = 5.74$, 6.00 and 6.26 [5], as well as the two UKQCD points at $\beta = 6.2$, with improved Wilson action [6]. The line represents a linear fit to all our nine points, leading to a first estimate for $\Delta_{\Lambda_b}$; the cross marks the final estimate, keeping into account finite $a$ effects. At $\frac{1}{M_P} = 0$, we plotted the three static estimates for $\Delta_{\Lambda_b}$, given, from the highest to the lowest, by references [4], [3] and [7].

## 6. Final results and conclusions

Using the experimental value for the $B-$meson mass $M_B = 5.27$ GeV, we derive from (5) the estimate:

$$M_{\Lambda_b} = 5.728 \pm 0.144 \pm 0.018 \text{ GeV}. \qquad (6)$$

Following the same procedure, we obtain for the $\Xi_b$ baryon the result $M_{\Xi_b} = 10.45 \pm 0.17$ GeV.

For the $\Lambda_b$, comparisons with other lattice estimates are possible, and they are summarized in fig.1. Looking at this picture, we note first that the Wilson fermion results from UKQCD, produced using an improved Wilson action at $\beta = 6.2$ [6], fall nicely on the same line as our data, confirming that the finite $a$ effects on this quantity are rather small, as seen also from the fact that our two estimates with and without finite $a$ effects practically coincide. As for the static points, the most recent one by UKQCD [7], obtained using an improved action, is in good agreement with the non-static results, whereas the previous (rough) estimates [4,3] seemed slightly too high: this might have been due to contamination form higher states, and finite $a$ effects. Finally, the comparison with experimental data is also satisfying: the extrapolations agree with the value measured by the UA1 collaboration [10] $M_{\Lambda_b} = 5.640 \pm 0.050 \pm 0.030$ GeV.

In conclusion, our final results are in good agreement with other lattice estimates and with the experimental values. Consistence between static and non-static results is also achieved thanks to the new UKQCD static estimate. In general, the picture emerging from measures and lattice estimates of $M_\Lambda - M_B$ is a consistent one, and contains good indications that, for this quantity, finite $a$ effects are small and under control, confirming that lattice computations of mass splittings are a reliable method for estimating baryon masses.

## REFERENCES

1. F. Butler et al., IBM-Report 1994 (HEP-LAT-9405003).
2. C.Bernard in *Lattice 93*, Nucl.Phys. B(Proc. Suppl.)34(1994)47; R.Sommer, these proc.
3. C.Alexandrou et al., Nucl. Phys. B414 (1994) 815.
4. M. Bochicchio et al., Nucl. Phys. B372 (1992) 403.
5. C.Alexandrou et al., Phys.Lett B337 (1994) 340.
6. UKQCD Collab., N.Stella, these proc.
7. UKQCD Collab., A.Ewing, these proc.
8. S.Güsken, in *Lattice 89*, Nucl. Phys. B(Proc. Suppl.) 17 (1990) 361, A.Duncan et al., Nucl. Phys. B(proc. Supp.) 30(1993)441, UKQCD Collab., S.Collins, Nucl. Phys. B(Proc. Supp.)30 (1993) 393.
9. C.Alexandrou et al., Z. Phys. C 62 (1994) 659.
10. UA1 Collab., Phys. Lett. B273(1991)540.